# High School Computer Science Participation: A 6-Year Enrollment Study

Cynthia L. Blitz[1], David J. Amiel[1] iD, & Teresa G. Duncan[2] iD

[1] CENTER FOR EFFECTIVE SCHOOL PRACTICES, RUTGERS UNIVERSITY
[2] DEACON HILL RESEARCH ASSOCIATES, LLC.

High-quality instruction in computer science (CS) is increasingly recognized as a vital component of K-12 education, equipping students with foundational skills that are critical for success in an increasingly technology-driven world. High school (HS) CS courses serve as a key entry point in this educational pipeline, shaping students' understanding and identity in the discipline, introducing them to key concepts, developing problem-solving and computational thinking techniques, and building awareness of educational and career opportunities in the field (Armoni & Gal-Ezer, 2022). Despite growing recognition of CS education's importance, and an increasing number of U.S. high schools offering CS courses, overall participation rates remain modest, raising important questions about engagement and enrollment (Code.org et al., 2024).

As computing, artificial intelligence, and other emerging technologies continue to influence various industries, early exposure to CS concepts and a computational mindset (Wing, 2006) help prepare students for a wide range of career paths. To inform policies and instructional strategies that encourage participation and achievement in CS, it is important to examine enrollment trends in HS CS in a variety of ways (Chan et al., 2022; Freeman et al., 2024). Doing so highlights the extent to which CS education is reaching students, engaging and retaining them, and encouraging them to remain on the CS pipeline. Here, we share one such analysis, which examines HS CS participation among 7 schools in the Northeastern United States over 6 academic years.

**Methodology**. This study utilizes de-identified student-level administrative data from seven public high schools, each located in large suburban districts. Data were collected from the 2018-2019 through the 2023-2024 academic years, encompassing a total of 329,961 high school course enrollments from 15,861 unique students in grades 9-12. Courses were classified as "CS" or "non CS" using the School Courses for the Exchange of Data (SCED) system (National Forum on Education Statistics, 2023). Courses assigned SCED code 10 (Information Technology) or 21 (Engineering and Technology) were considered CS courses, and classifications were manually verified. For each academic year, we calculated the percentage of students who enrolled in at least one CS course. Additionally, we conducted separate analyses for foundational CS courses (e.g., introductory programming, robotics, computer applications, or design) and advanced CS courses (e.g., AP CS courses and higher-level programming and robotics courses).

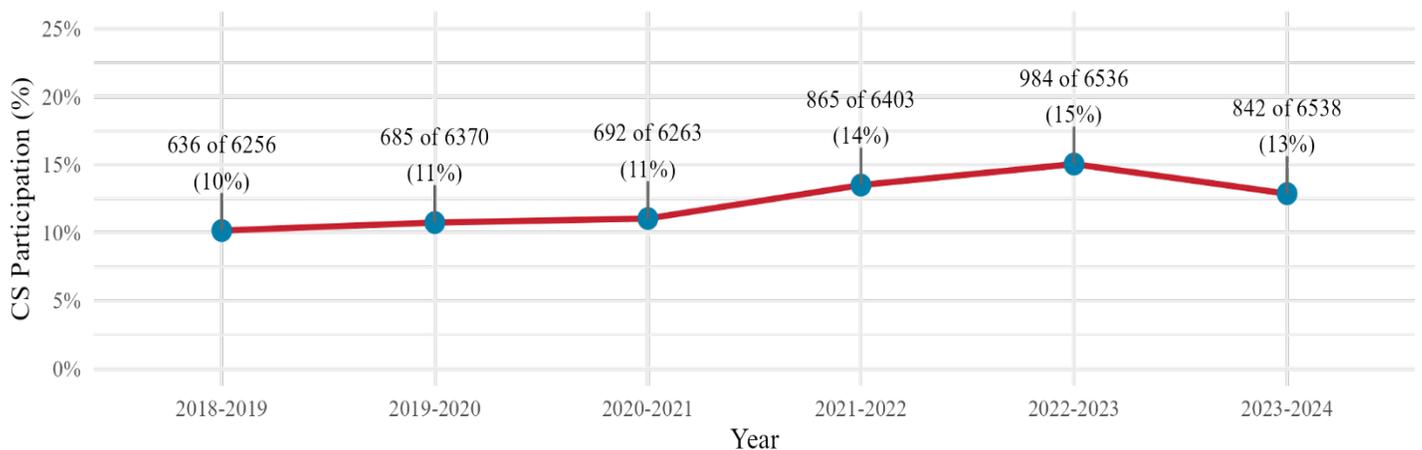

*Figure 1.* HS Computer Science Participation Over Time

Research Brief

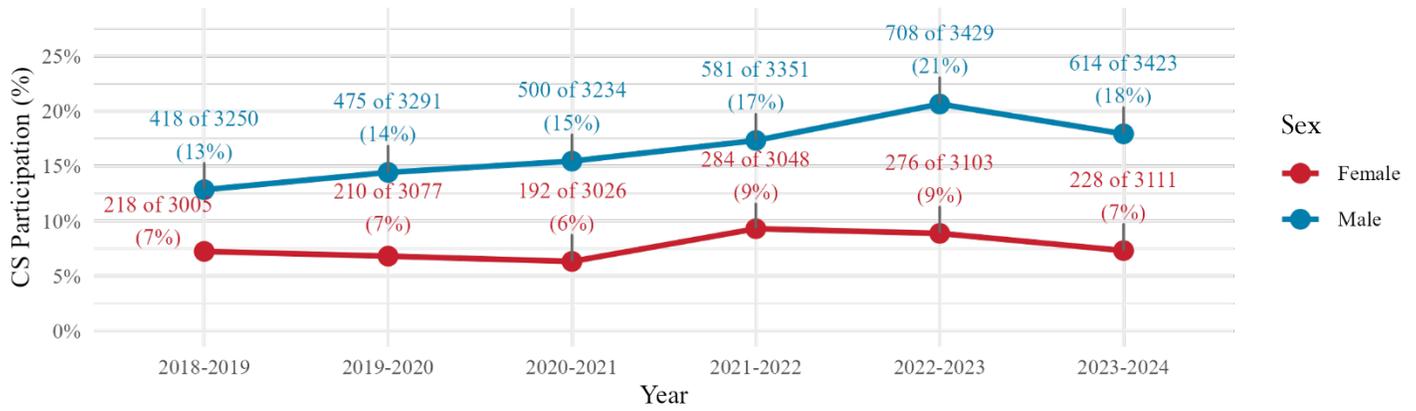

*Figure 2.* HS Computer Science Participation Over Time by Sex

**Findings: Overall CS Participation**. As shown in *Figure 1*, overall participation in HS CS courses remained relatively steady over time. In the 2018-2019 academic year, 10 % of students (636 of 6,256) enrolled in at least one CS course. Participation remained stable until 2020-2021, followed by a moderate increase to 15% by 2022-2023. However, participation declined to 13% in the 2023-2024 academic year. While it is uncertain whether this decline represents a larger trend, further monitoring is warranted.

**Findings: Participation by Sex**. Grouping students by sex reveals a clear and consistent difference between males' and females' participation in CS courses, as shown in *Figure 2*. Across all 6 academic years, females participated in HS CS courses at a lower rate than their male peers. Additionally, examining the data this way reveals that male participation in CS steadily increased from the 2018-2019 academic year, where 13% took at least one CS course, through the 2022-2023 academic year, where 21% of males took a course. Female CS participation, on the other hand, decreased in each successive academic year since 2018, apart from the 2020-2021 to 2021-2022 academic years. When examined together, CS participation appears relatively steady, though looking at male and female participation reveals diverging participation rates over time, which appear flat when averaged together, as in Figure 1. The rate of female participation in HS CS is the same in the 2023-2024 academic year as it was in the 2018-2019 academic year (7%), whereas the male participation increased by 5% in this period (from 13% to 18%).

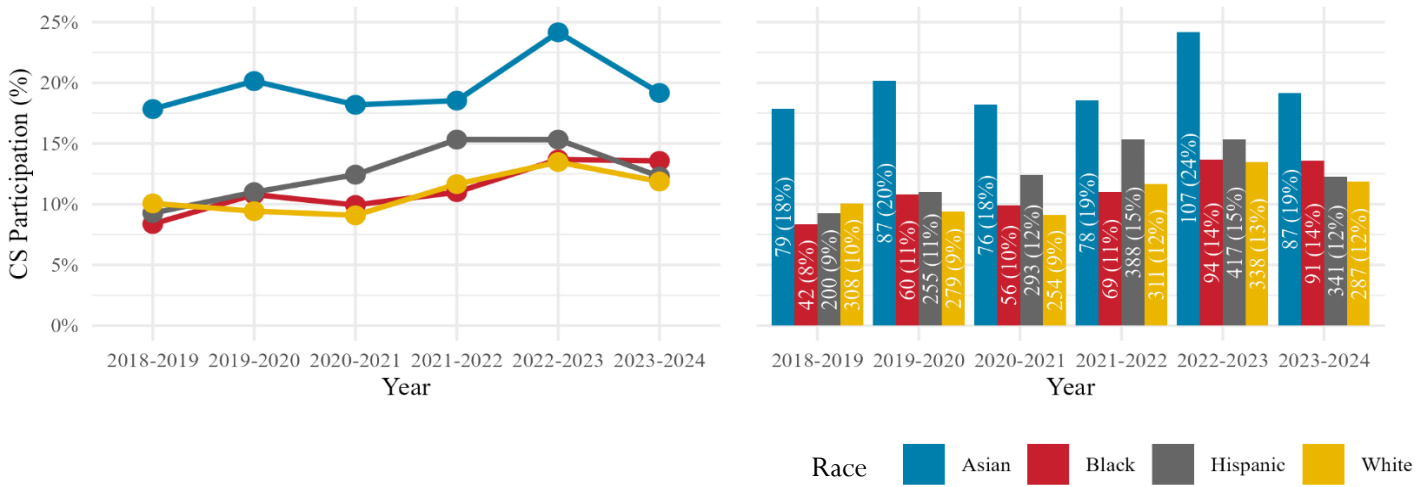

*Figure 3.* HS Computer Science Participation Over Time by Race

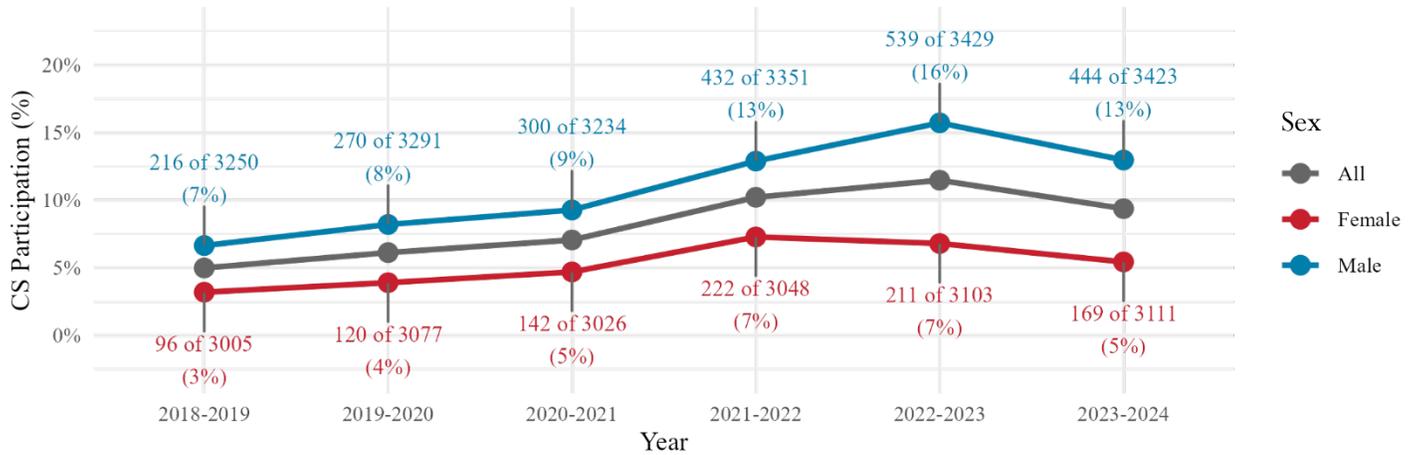

*Figure 4.* HS Foundational Computer Science Participation Over Time, Overall & By Sex

**Findings: Participation by Race**. Enrollment data indicate variation in CS participation across racial groups, as shown in *Figure 3*. Asian students consistently participated in CS courses at higher rates than their peers, with participation reaching a six-year high of 24% in the 2022-2023 academic year before declining to 19% in 2023-2024. However, it is more difficult to distinguish clear differences in participation among Black, Hispanic, and White students. Hispanic student participation showed notable growth, peaking at 15% in 2021-2022, surpassing Black and White student participation rates. However, this increase was not sustained, as participation declined in the following years. Black and White student participation remained relatively stable, generally fluctuating between 10-12% over the study period. Racial differences in CS participation are often more difficult to interpret given many factors at play, such as each racial group's share of the overall student population, which varies greatly across the participating schools. The trends found in this data highlight the need for continued research into the factors influencing CS enrollment across different student populations.

**Findings: Foundational CS Participation**. Rates of participation in foundational CS courses appear very similar to participation in CS courses overall. This can be explained in part by pre-requisite structures in schools where a foundational course may be required before taking an advanced course, but this is not always the case (as a primary example, AP CS Principles, an advanced course, is frequently offered as the first CS course in HS). From the 2018-2019 academic year through the 2021-2022 academic year, there was a slight upward trajectory in foundational CS participation, as shown in *Figure 4*, and male and female participation rates, although different, moved in parallel. From the 2021-2022 through the 2023-2024 academic years, male and female trajectories diverged, with the gap between male and female participation reaching a high in 2022-2023, when 16% of males took a foundational CS course, compared to only 7% of females. Importantly, although female participation in all CS courses declined during the first 3 years, female participation in foundational courses rose slightly during this period.

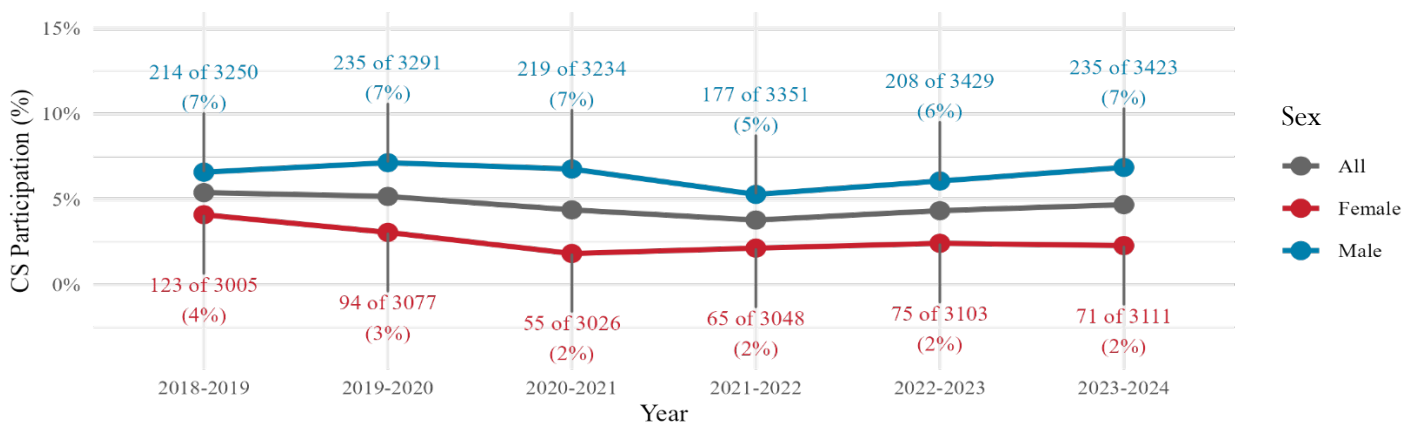

*Figure 5.* HS Advanced Computer Science Participation Over Time, Overall & By Sex

**Findings: Advanced CS Participation.** Generally, participation in advanced computer science courses is lower than foundational CS. Females also participate in these courses at consistently lower levels than their male peers, as seen in *Figure 5*. For instance, in the 2020-2021 academic year, only 55 of 3,026 females, 2%, took an advanced CS course; in the same year, 219 of 3,234 males, 7%, did so. However, fluctuations in growing between males and females in recent years, and overall participation in advanced CS does not drop in the 2023-2024 academic year. The participation in advanced CS is also lower because there are fewer advanced CS options available to students; given that advanced CS course-taking in HS represents a small piece of this picture (and one where the data is sensitive to differences in access across schools), understanding advanced CS participation requires further investigation in more focused ways.

**Conclusion.** While access to HS CS courses has expanded, overall participation rates have remained relatively stable over the past six years. Despite modest growth in participation through 2022-2023, the decline observed in 2023-2024 highlights the need for further investigation into factors influencing enrollment trends. The data suggest that while some student groups have experienced incremental increases in participation, overall trends remain largely unchanged. Efforts to increase participation in HS CS courses must go beyond simply providing access; they should also focus on fostering engagement, addressing barriers to enrollment, and ensuring that students recognize the value of computing skills in various career pathways. As technology continues to shape the future of work, sustained efforts are needed to ensure that HS CS education effectively prepares students for opportunities in computing and beyond. There most certainly remains work to be done to break away from a slow climb and set CS participation on a trajectory of meaningful growth.

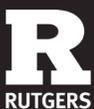

## The Center for Effective School Practices
### Excellence & Integrity in Research & Evaluation

Under the executive leadership of Dr. Cynthia L. Blitz, research professor at the Graduate School of Education, the Center for Effective School Practices (CESP) is a practice-focused unit at Rutgers University dedicated to excellence and integrity in research and evaluation. Rutgers CESP regularly engages in and mediates collaborations among public and private school districts in the tri-state area, institutions of higher education, local, state, and federal government agencies, community organizations, and industry partners to generate and implement practitioner-relevant best-practices in education.

The Rutgers EIR: Extending the CS Pipeline is a research grant dedicated to partnering with middle schools to enhance the rigor and relevance of computer science and related instruction. The project is funded by the United States Department of Education Office of Elementary & Secondary Instruction under Educational Innovation & Research (EIR) Award #S411C200084. This study was reviewed and ap-proved by the Institutional Review Board of Rutgers University, New Brunswick (study 2020003169, approved 5/28/2021). The datasets used in this study are not publicly available, as their access was provided by public educational agencies with researchers through board-approved, time-bound data sharing agreements.

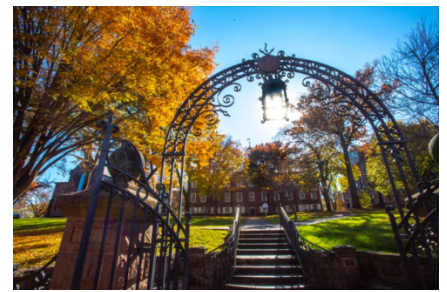